\documentclass[a4paper,twocolumn,11pt,unpublished]{quantumarticle}
\pdfoutput=1

\usepackage[utf8]{inputenc}
\usepackage[english]{babel}
\usepackage[T1]{fontenc}
\usepackage{amsmath}
\usepackage{hyperref}
\usepackage{cleveref}
\usepackage{appendix}
\usepackage[numbers,sort&compress]{natbib}
\usepackage{placeins}

\usepackage{booktabs}

\newcommand{\new}[1]{\textcolor{black}{#1}}

\begin{document}

\title{Optimizing ZX-Diagrams with Deep Reinforcement Learning}
\author{Maximilian N\"agele\textsuperscript{\textdagger,}}
\affiliation{Max Planck Institute for the Science of Light, Staudtstraße 2, 91058 Erlangen, Germany}
\affiliation{Physics Department, Friedrich-Alexander-Universit\"at Erlangen-N\"urnberg, 91058 Erlangen, Germany}
\orcid{0000-0001-6382-2077}
\author{Florian Marquardt}
\affiliation{Max Planck Institute for the Science of Light, Staudtstraße 2, 91058 Erlangen, Germany}
\affiliation{Physics Department, Friedrich-Alexander-Universit\"at Erlangen-N\"urnberg, 91058 Erlangen, Germany}
\date{\textsuperscript{\textdagger}
Author to whom any correspondence should be addressed. E-mail:\,\href{mailto:maximilian.naegele@mpl.mpg.de}{maximilian.naegele@mpl.mpg.de}}
\orcid{0000-0003-4566-1753}

\begin{abstract}
    ZX-diagrams are a powerful graphical language for the description of quantum processes with applications in fundamental quantum mechanics, quantum circuit optimization, tensor network simulation, and many more. The utility of ZX-diagrams relies on a set of local transformation rules that can be applied to them without changing the underlying quantum process they describe. These rules can be exploited to optimize the structure of ZX-diagrams for a range of applications. However, finding an optimal sequence of transformation rules is generally an open problem. In this work, we bring together ZX-diagrams with reinforcement learning, a machine learning technique designed to discover an optimal sequence of actions in a decision-making problem and show that a trained reinforcement learning agent can significantly outperform other optimization techniques like a greedy strategy, simulated annealing, and \new{state-of-the-art hand-crafted algorithms}. The use of graph neural networks to encode the policy of the agent enables generalization to diagrams much bigger than seen during the training phase.
\end{abstract}

\maketitle

\section{Introduction}
ZX-calculus is a diagrammatic language for the representation of quantum processes as graphs equipped with a set of local transformation rules. Due to the utility of these transformation rules, ZX-calculus has been applied to a wide range of problems ranging from fundamental quantum mechanics\,\cite{Coecke2017} over the description of measurement-based quantum computing\,\cite{Duncan2012} and analyzing variational quantum circuits\,\cite{Martin2023} to quantum error correction\,\cite{Chancellor2023, Garvie2017}. In particular, ZX-calculus has proven a promising candidate for speeding up tensor network simulations\,\cite{Cam2023} and quantum circuit optimization\,\cite{Duncan2020, Staudacher2022, Gogioso2022, Kissinger2020, Winderl2023}.
However, finding the optimal sequence of transformation rules to achieve a given task is often a non-trivial task. 
Therefore, we bring together ZX-diagrams with reinforcement learning (RL), a machine learning technique where an agent iteratively interacts with an environment to learn a policy predicting an optimal sequence of actions. RL has been successfully applied to various domains such as game-playing\,\cite{Mnih2013, Silver2018}, robotics\,\cite{Kober2013, Andrychowicz2020}, quantum chemistry\,\cite{You2018, Shi2020}, and problems in quantum computing like quantum error correction\,\cite{Foesel2018, Olle2023, Sweke2021}, quantum control\,\cite{Baum2021, Reuer2022}, and circuit optimization\,\cite{Fosel2021, li2023}.
\new{For optimizing ZX-diagrams, RL has two advantages over other machine learning methods: First, it doesn't require a training set of optimized diagrams, which is not available in our case. Second, by iteratively rewriting the diagram using RL instead of directly generating the optimized diagram in a single shot using supervised methods, we can verify the equivalence of the unoptimized and its corresponding optimized diagram, which is otherwise exponentially hard in the number of inputs and outputs of the diagram. Ensuring this equivalence is crucial for the task we set out to solve.}

To capitalize on the graph structure of ZX-diagrams, we encode the policy of our reinforcement learning agent as a graph neural network (GNN)\,\cite{Zhou2020}. 
\new{As a proof-of-principle application, we train the agent to reduce the number of nodes in random ZX-diagrams for the following reasons: 
\begin{itemize}
    \item It is intuitive for humans to understand, and good heuristic algorithms are available for benchmarking.
    \item It allows the use of a complete set of transformation rules, making it a more fundamental task as opposed to more specialized problems such as quantum circuit optimization\,\cite{Kissinger2020}.
    \item It is computationally challenging, and, in fact, QMA-complete\,\cite{JANZING2005}.
\end{itemize}
We show that the agent learns a non-trivial strategy outperforming a custom greedy strategy, simulated annealing, and handcrafted state-of-the-art ZX-diagram optimizers originally developed for quantum circuit optimization\,\cite{Kissinger2020, Duncan2020}.} Moreover, the agent's policy generalizes well to diagrams much larger than seen during training.
Our work lays the foundation for applying the combination of RL and ZX-calculus to a broad range of tasks like minimizing the gate count of quantum circuits, \new{speeding up tensor network simulations, or checking quantum circuit equivalence, by changing the optimization goal of the agent in future work.}

\begin{figure*}[t]
    \centering
    \includegraphics{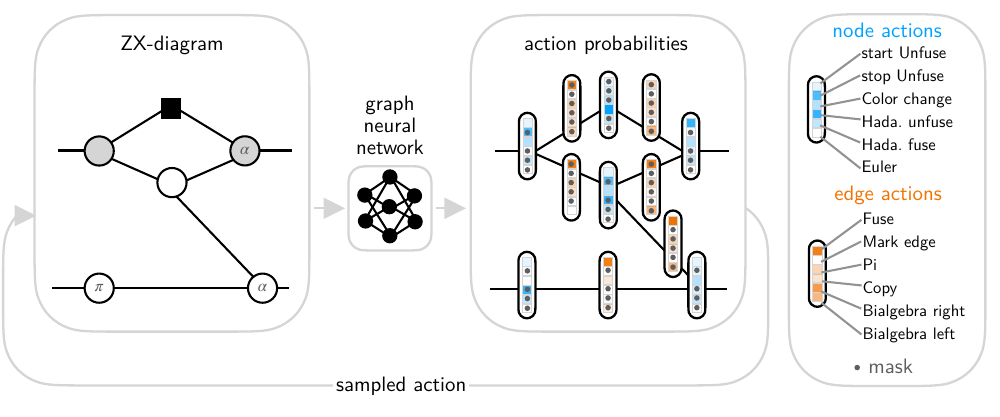}
    \caption{Schematic of the optimization loop. At each step, the reinforcement learning agent is provided with a ZX-diagram in the form of a graph. The agent then uses a graph neural network to suggest action probabilities of local graph transformations (color-coded), which act on either a unique edge (orange) or node (blue). Finally, an action is sampled from this probability distribution and applied to the diagram. In total, there are 6 separate actions per node and edge, some of which are not allowed in their local environment and, therefore, masked (grey dots). For a definition of the graph transformations see \Cref{fig:fig_transfo}.}
    \label{fig:fig_rl}
\end{figure*}

\section{ZX-diagrams}

A ZX-diagram is a graph representation of a quantum process defined by an arbitrary complex matrix of size $2^k \times 2^j$, where $j$ is the number of ingoing and $k$ the number of outgoing edges of the diagram. For example, we can represent the following matrix either as a quantum circuit consisting of single qubit X- and Z-gates and a CNOT gate or as a ZX-diagram according to
\begin{equation}
    \includegraphics{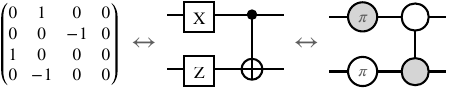}.
\end{equation}
The central building blocks of ZX-diagrams are Z-spiders (white) and X-spiders (grey) defined as
\begin{equation}\label{eq:spider_def}
\includegraphics{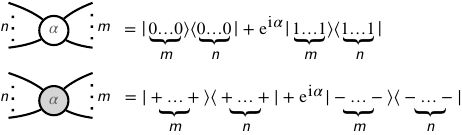},
\end{equation}
where $|1/0\rangle$ ($|+/-\rangle$) are the eigenvectors of the Pauli-Z (Pauli-X) matrix, $\alpha$ is an angle, and $n$ and $m$ are non-negative integers specifying the amount of input and output edges of the spider.
\new{In addition to unitary operations, ZX-diagrams can also contain states and post-selected measurements. Therefore, a translation from quantum circuits into ZX-diagrams is straightforward, while the reverse is not always possible.}
While multiple different ZX-diagrams can describe the same underlying matrix, they can be transformed into each other by the set of local transformation rules depicted in \Cref{fig:fig_transfo}, which are correct up to a non-zero scalar factor\,\cite{Vilmart2019}. These rules also imply that multiple edges connecting spiders of the same color can be reduced to just one edge and multiple edges between spiders of differing colors can be taken modulo two. 
Therefore, and due to the inherent symmetries of the Z- and X-spiders, ZX-diagrams can be regarded as simple graphs\,\cite{Wetering2020}. For a more detailed introduction of ZX-diagrams, see \Cref{sec:details_zx}.

\section{Optimization of ZX-diagrams as a reinforcement learning problem}
\begin{figure*}[t]
    \centering   
    \includegraphics{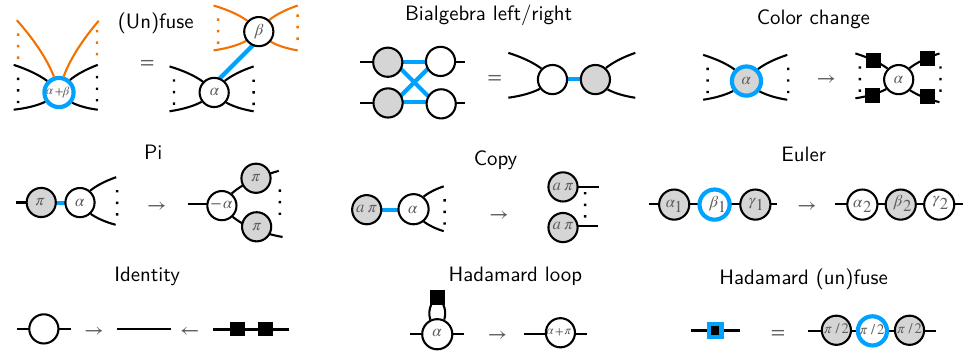}
    \caption{Encoding of the local transformation rules of ZX-diagrams as actions of a reinforcement learning agent. Blue colors indicate the encoding as an action of the agent acting on either an edge or a node. Some transformations are implemented in both directions as separate actions of the reinforcement learning agent (equal signs), while some are only implemented in one direction (arrows).
    Three dots stand for zero or more edges. Each rule also holds with the spiders' colors inverted and in both directions. Black squares represent a Hadamard gate as defined by the \emph{Hadamard fuse} transformation. 
    During the \emph{Unfuse} transformation, a spider is split into two by arbitrarily splitting up its angle between the two resulting spiders, connecting them with a new edge, and transferring a subset of the originally connected edges (orange) to the new spider. In the \emph{Copy}  transformation, $a\in {0,1}$. In the \emph{Euler} transformation, $\alpha_1/\beta_1/\gamma_1$ are related to $\alpha_2/\beta_2/\gamma_2$ by trigonometric functions as defined in\,\cite{Vilmart2019}.
    }\label{fig:fig_transfo}
\end{figure*}

Reinforcement learning (RL) is a machine learning technique where an agent recursively interacts with an environment during a trajectory comprising multiple steps. At each step $t$, the agent uses its policy to select an action (in our case a graph transformation) based on an observation describing the environment's state (in our case a ZX-diagram) as depicted in \Cref{fig:fig_rl}. 
This action then modifies the state of the environment and a new observation and a numerical value, the reward $r_t$ (in our case the difference in node number between the old and new diagram), is supplied to the agent. This scheme continues until the environment terminates the trajectory after a fixed amount of steps or the agent chooses a special \emph{Stop} action. 
The agent is trained by repeating two phases: During the sampling phase the agent interacts with the environment for a fixed amount of steps. Then, during the training phase, the agent's policy is updated to maximize the expected cumulative reward over a complete trajectory $\langle\sum_t \gamma^t r_t\rangle$, where $\gamma$ is the discount factor\,\cite{Sutton2018}. 
To enable the use of graph neural networks to encode the agent's policy, we use a custom implementation of a state-of-the-art reinforcement learning algorithm named Proximal Policy Optimization (PPO)\,\cite{Schulman2017} to train the agent (for details on the algorithm and ablation studies of its features see \Cref{sec:details_ppo}).

Each of the transformation rules of ZX-diagrams acts only on the local neighborhood of an edge or node. We can, therefore, identify each possible action of the RL agent with either a unique node or a unique edge as indicated by the blue lines in \Cref{fig:fig_transfo}. The agent's policy then predicts the unnormalized log-likelihood for each of the possible actions. By normalizing over the whole diagram, we build a probability distribution from which we sample an action that is applied to the diagram (see \Cref{fig:fig_rl}).

Some of the transformations are symmetric and only implemented in one direction (arrows) resulting in one action. For example, the \emph{Color change} transformation changes the color of a spider by inserting Hadamards on all connected edges. Because of the \emph{Identity} transformation, an implementation in the other direction would be redundant.
Other transformations need to be implemented in both directions (equal signs). For example, the \emph{Hadamard (un)fuse} transformation either fuses three spiders into a single Hadamard or splits up a Hadamard into three spiders and needs to be implemented as two separate actions.

The \emph{Unfuse} transformation is especially challenging to implement since it requires choosing a subset of the edges connected to the selected node. As the number of edges connected to a spider is in principle unbounded, defining multiple \emph{Unfuse} actions, each corresponding to separating the spider with a specific division of its edge set is not feasible.
As a solution, we split up the \emph{Unfuse} transformation into multiple consecutive actions of three types. First, the \emph{start Unfuse} action selects a spider. After that, the selected spider is marked by a new node feature and the agent can only select one of two actions at each step: It can iteratively either use the \emph{Mark edge} action on an edge connected to the selected spider or select the \emph{stop Unfuse} action. Once the \emph{stop Unfuse} action is selected, the spider is split and all previously marked edges (orange edges in \Cref{fig:fig_transfo}) are moved to the newly created spider. The angle remains fully at the original spider. To also split the angle between both spiders, as an extension, multiple different \emph{stop Unfuse} actions could be defined, each standing for a different angle of the newly created spider.
Due to the symmetry of the \emph{Bialgebra right} transformation, it can not be identified with a single unique edge. Instead, it is applied if the agent selects one of the corresponding $4$ colored edges.
Due to its potentially global properties, the \emph{Copy} transformation is only implemented in one direction. In principle, the other direction could be implemented similarly as the \emph{Unfuse} action by first iteratively marking all participating nodes.

In total, the agent can choose from 6 different actions for each node and each edge of the considered diagram. Additionally, the agent can always select a global \emph{Stop} action to end a trajectory if it expects that it can't optimize the diagram any further. To enable more efficient training, we mask actions that are not allowed in their local environment by setting their probability to $0$.
Finally, after each step, possible \emph{Identity} and \emph{Hadamard loop} transformations are applied automatically and redundant edges are removed. We also delete parts of the ZX-diagram that are disconnected from all ingoing and outgoing edges, as they correspond to simple scalar factors.

To encode ZX-diagrams as observations supplied to the agent, we represent them as undirected graphs with one-hot encoded node features. Each node has a color feature that can either be Z-spider, X-spider, Hadamard, Input, or Output  and is, therefore, represented as a $5$ dimensional vector. For example, the color feature of an X-spider would be \mbox{$[0,1,0,0,0]$}.
The Input and Output nodes are used to define the ingoing and outgoing edges of the diagram by being connected to their otherwise open end. Additionally, each node has an angle feature that can either be an unspecified placeholder angle $\alpha$, multiples of $\pi/2$, or specify that the node is not a spider and doesn't have an angle. The angle feature, therefore, is a $6$ dimensional vector. The discrete multiples of $\pi/2$ are necessary to evaluate the possibility of the transformation rules depicted in \Cref{fig:fig_transfo}. Finally, each node has a binary feature indicating whether the node has been marked by the \emph{start Unfuse} action. 
The complete feature vector of each node $n$, $x_{0}^{n}$, given to the agent's policy is then all of its features concatenated into a single vector, resulting in a $12$ dimensional vector. The feature vector $e_{0}^{(n,m)}$ of the edge connecting node $n$ and $m$ contains just a single number that is $0$ if the edge has not been marked by the \emph{Mark edge} action and $1$ otherwise.

Finally, a vital part of every RL algorithm is the definition of the reward of the agent as it determines the optimization goal. To demonstrate the utility of our algorithm we choose a reward that is computationally cheap to evaluate and intuitive to understand for humans but still requires non-trivial strategies to maximize: The difference in node number of the diagram before and after action application. As the sum of those differences corresponds to the total change in node number, the agent tries to minimize the total amount of nodes in the diagram at the end of the trajectory. \new{Optimally reducing the node number of ZX-diagrams includes checking whether a given ZX-diagram corresponds, up to a global phase, to the identity operation which is a QMA-complete problem\,\cite{JANZING2005}.}

\section{Neural network architecture}

\begin{figure*}[h!t]
    \centering
    \includegraphics{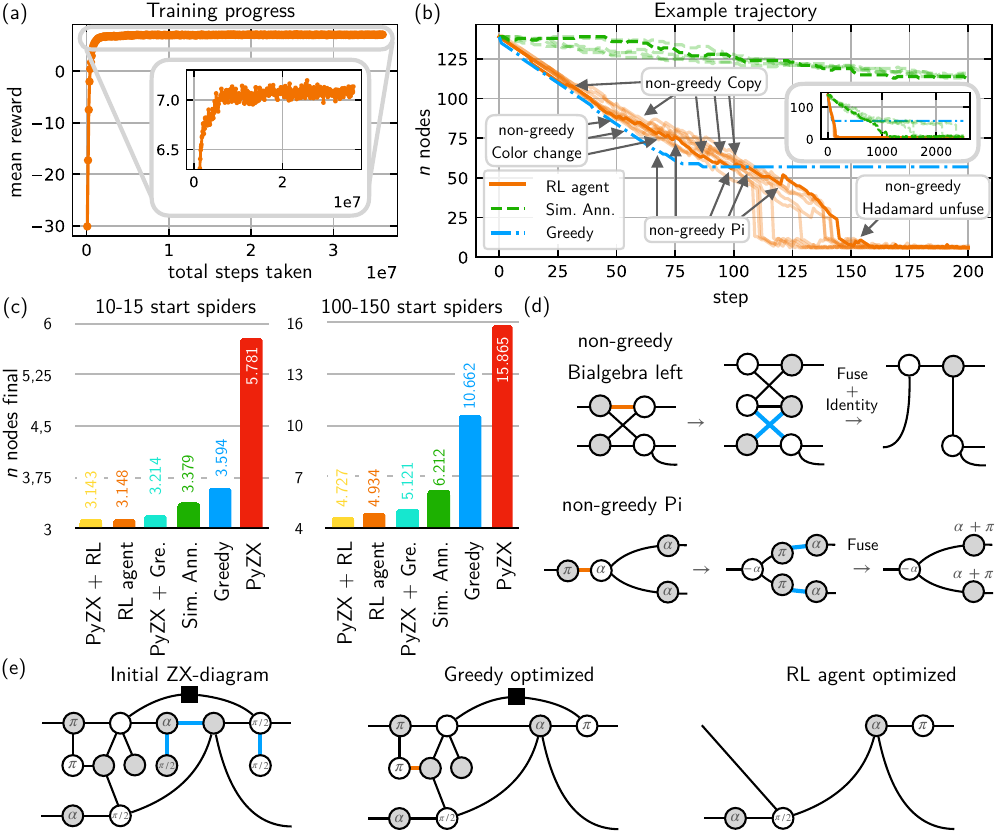}
    \caption{Results. (a)~Training progress as the agent is trained to reduce the node number in random ZX-diagrams. Mean cumulative reward of the agent per trajectory against total steps taken in the environment. (b)~Optimization of an example ZX-diagram ten times larger than the RL agent's training diagrams. Number of nodes in the ZX-diagram against each action taken for the RL agent (orange), a greedy strategy (blue), and simulated annealing (green). For the RL agent and simulated annealing, multiple trajectories are plotted (transparent). The RL agent and simulated annealing significantly outperform the greedy strategy in terms of cumulative reward with the RL agent requiring an order of magnitude less steps than simulated annealing (inlay). Actions taken by the RL agent that intermittently increase the node number (i.e.\ non-greedy actions) are indicated by arrows. (c)~Average number of nodes after optimization of $1000$ ZX-diagrams with $10$-$15$ initial spiders (left), which is the size the agent was trained on, and $100$-$150$ initial spiders (right). \new{We compare the RL agent (orange) to simulated annealing (green), a custom greedy strategy (blue), the \emph{full\textunderscore reduce} function of the \texttt{PyZX} software package (red), a combination of \emph{full\textunderscore reduce} and the greedy strategy (turquoise), and a combination of \emph{full\textunderscore reduce} and the RL agent (yellow).}
    Hyperparameters for simulated annealing are optimized to give good performance on two example diagrams and then kept fixed for all diagrams. \new{The RL agent is outperforming the other strategies.} (d)~Two examples of non-greedy actions learned by the agent (orange lines), that lead to a positive cumulative reward by consecutive \emph{Fuse} actions (blue lines). (e)~Example ZX-diagram sampled from the agent's training set. The greedy strategy can reduce the node number by applying $3$ \emph{Fuse} actions (blue lines) while the agent further optimizes the diagram beginning with a non-greedy \emph{Pi} action (orange line).}
    \label{fig:fig_result}
\end{figure*}

The use of a graph neural network (GNN) to encode the agent's policy has several advantages: As we suppose the ideal policy depends only on the local structure of the ZX-diagram, we expect the GNN to train more efficiently and generalize better to unseen diagrams than other neural network architectures. Also, unlike a dense neural network, the GNN can handle any size of input data. Therefore, it can be efficiently trained on relatively small diagrams and later straightforwardly applied to much bigger diagrams.

 As an input, the GNN directly takes the graph representation of a ZX-diagram. 
 First, $6$ message-passing layers\,\cite{Gilmer2017} are applied to the graph. At each layer $i$, the node feature vectors $x_i^{n}$ are updated according to 
\begin{equation}
    x_{i+1}^{n} = \phi_i\left( x_{i}^{n}, \underset{x_{i}^{m} \in \mathcal{N}_n}{\sum}\left[\psi_i\right(x_{i}^{n}, x_{i}^{m}, e_{i}^{(n,m)}\left)\right]\right),
\end{equation}

where $\mathcal{N}_n$ are the nearest neighbors of node $n$, and $\phi_i$ and $\psi_i$ are single dense neural network layers. 
We also update the edge feature vectors at each layer according to
\begin{equation}
    e_{i+1}^{(n,m)} = \theta_i\left(e_{i}^{(n,m)}, x_{i}^{n}, x_{i}^{m} \right),
\end{equation}
where $\theta_i$ is also a single dense neural network layer.
After the message-passing layers, we apply the multi-layer perceptron $\chi_\mathrm{node}\left(x_\mathrm{f}^n\right)$ to the final features of each node $x_\mathrm{f}^n$ and the multi-layer perceptron $\chi_\mathrm{edge}\left(e_\mathrm{f}^{(n,m)}\right)$ to the final features of each edge $e_\mathrm{f}^{(n,m)}$. The networks $\chi_\mathrm{node}$ and $\chi_\mathrm{edge}$ have $6$ output neurons each which are interpreted as the unnormalized log-probabilities of the possible actions (see \Cref{fig:fig_rl}).

As the \emph{Stop} action of the agent depends not only on the local structure of the graph but also on global features, we treat it differently from the other actions by computing its unnormalized log-probability according to
\begin{equation}
    p_\mathrm{s} = \chi_\mathrm{stop}\left(C,  \underset{n}{\mathrm{MEAN}}\left(x_\mathrm{f}^n\right), \underset{(n,m)}{\mathrm{MEAN}}\left(e_\mathrm{f}^{(n,m)}\right)\right),
\end{equation}
where $\chi_\mathrm{stop}$ is a multi-layer perceptron, the $\mathrm{MEAN}$ functions are taken over the final node/edge features, and $C$ is a vector containing global information about the amount of each node type, edges and allowed actions.

For an efficient implementation of the GNN, we use the TensorFlow-GNN software package\,\cite{Ferludin2022} with custom layers to handle undirected edges. For a graphical representation of the network and further details on the network architecture and implementation see \Cref{sec:details_network}.

\section{Results}

\subsection{Training}

We train the agent to reduce the node number in randomly sampled ZX-diagrams with $10$-$15$ initial spiders (for details on the diagram sampling see \Cref{sec:details_sampled}). The agent is trained for a total of $36*10^6$ total actions. However, it already reaches its optimal performance around $9*10^6$ actions as shown in \Cref{fig:fig_result}\,(a).
To evaluate the trained agent, we sample $1000$ new ZX-diagrams of the same size as the training set and optimize them for $200$ steps. We then calculate the average of the minimum number of nodes found during optimization which is significantly lower than the number of initial nodes. Next, we want to answer the question of whether the learned policy can straightforwardly be applied to larger diagrams by repeating the same evaluation on ZX-diagrams with $100$-$150$ initial spiders. Even though the agent was only trained on diagrams an order of magnitude smaller, it can reduce the number of nodes in the diagram substantially, thereby highlighting the powerful generalization ability of GNNs [see \Cref{fig:fig_result}\,(c)]. 
To demonstrate the need for non-trivial strategies of the trained agent to achieve these results, we show two selected actions that initially increase the spider number but later lead to an overall positive cumulative reward in \Cref{fig:fig_result}\,(d).

Training the agent takes around $41$ hours on a single compute node with $32$ CPUs and $2$ GPUs. We run multiple environments in parallel on the CPUs during the sampling phase and train the agent distributed on both GPUs. The implementation of the algorithm could directly take advantage of larger compute nodes to speed up training time.

\subsection{Comparison with other techniques}

To better estimate the agent's performance, we compare it with \new{various other strategies}. 
The greedy strategy always selects the action with the highest possible reward as long as there are actions with a non-negative reward available. If there are multiple actions leading to the highest possible reward, the greedy strategy chooses randomly out of them. Simulated annealing is a probabilistic strategy for non-convex global optimization problems\,\cite{Henderson2003}. We optimize its hyperparameters, i.e.\ the start temperature and temperature annealing schedule, by hand on two example diagrams [used for \Cref{fig:fig_result}\,(b) and (e)] and then keep them fixed. For more details on the simulated annealing algorithm see \Cref{sec:details_sim_ann}.
\new{The PyZX strategy uses the most powerful routine of the \texttt{PyZX} software package, i.e.\ the \emph{full\textunderscore reduce} function, which is based on the circuit optimization algorithms presented in \cite{Duncan2020, kissinger2019}. Because the \emph{full\textunderscore reduce} function is used for circuit optimization, it uses an incomplete set of transformation rules and doesn't perform well on the node reduction task. Therefore, we also apply the greedy strategy (PyZX + Greedy) or the RL agent (PyZX + RL) after the \emph{full\textunderscore reduce} function.}

\new{To compare the different strategies, we evaluate them on the same two sets of $1000$ diagrams as the RL agent. 
The RL agent outperforms even the best non-RL strategy [see \Cref{fig:fig_result}\,(c)] without relying on human-designed task-specific algorithms, suggesting promising results when applied to related ZX-diagram optimization tasks in future work. The PyZX + RL strategy is slightly better than the pure RL agent, indicating that some of the composite rules used in the PyZX strategy are difficult for the agent to learn.}

The RL agent needs on average less than $4$\,s to simplify a diagram with $100$-$150$ initial spiders running on a single GPU and single CPU while simulated annealing with $20000$ steps needs over $36$\,s and the greedy strategy over $100$\,s, albeit running on only a CPU. \new{Both PyZX and PyZX + Greedy are significantly faster than the RL agent. However, the run-time of the PyZX and the RL approaches are expected to scale equally with the size of the diagram (see \Cref{sec:scaling}).}

\new{ZX-diagrams containing only Clifford spiders, i.e.\ spiders with phases, that are multiples of $\pi/2$, can be efficiently simulated classically\,\cite{Aaronson2004}. Therefore, it is often more important to reduce the number of non-Clifford spiders, i.e.\ the $\alpha$ spiders with an unspecified angle. Thus, we evaluate the performance of the same trained RL agent for reducing $\alpha$ spiders and find it outperforms the other strategies also on this task (see \Cref{tab:num_results}).}

\subsection{Analysis of learned policy}\label{sec:analysis}

\begin{figure*}[t]
    \centering
    \includegraphics{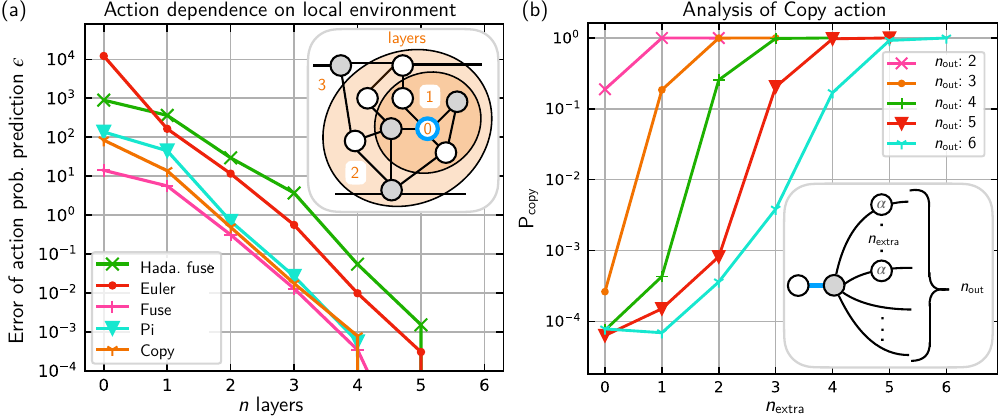}
    \caption{Analysis of learned policy. (a)~Action dependence on the local environment. 1000 actions of each type are sampled by the agent. Then, for each action and the diagram in which it was chosen, sub-diagrams are built up in layers around the node/edge identified with the action (see inlay). For each sub-diagram spanning only the nodes in a specific layer, we compute the agent's unnormalized probability of sampling the chosen action $P_\mathrm{layer}$ and compute the difference $\epsilon$ to its probability $P_\mathrm{complete}$ in the full diagram, where we define $\epsilon$ in \Cref{eq:error_dist}. We plot the average of this difference against the number of layers for $5$ action types.
    (b)~Probability of sampling the \emph{Copy} action on the blue edge in the diagram depicted in the inlay for multiple outputs of the diagram $n_\mathrm{out}$ and multiple additionally inserted spiders on the outputs $n_\mathrm{extra}$. The ideal strategy is to select the \emph{Copy} action for $n_\mathrm{out} - n_\mathrm{extra} \leq 2$. The agent approximately learns the ideal policy.
    }\label{fig:fig_interpret}
\end{figure*}
While deep neural networks have been successfully employed to solve a wide range of problems, they are often regarded as a 'black box method' due to difficulties in interpreting their learned strategies. However, it is in principle highly desirable to gain some insight into how the neural networks arrive at their predictions\,\cite{Lingfei2022}. For graph neural networks, an interesting quantity is how local their learned strategy is, i.e.\ how far away predictions on nodes or edges are influenced by the node and edge features of the diagram.
Therefore, we evaluate how far away from a chosen action the ZX-diagram still influences the agent's decision.

To this end, we optimize ZX-diagrams with the agent until $1000$ actions of each type are sampled. For each sampled action and the corresponding ZX-diagram, we then build up the diagram in layers around the node/edge identified with the action. Layer $n$ is defined as all nodes that can be reached in $n$ steps by traversing the diagram from the starting point.
For each layer $n$ and the corresponding sub-diagram spanning only nodes up to this layer, we compute the agent's unnormalized probability of sampling the original action $P_\mathrm{layer}$. We deliberately choose not to normalize the probabilities, as otherwise far away action probabilities would influence our results through the normalization constant even though no actual information traveled through the GNN. 

In \Cref{fig:fig_interpret}\,(a) we plot the average over the $1000$ sampled actions of the quantity $\epsilon$ which captures how different $P_\mathrm{{layer}}$ is from the unnormalized probability in the full diagram $P_\mathrm{{complete}}$. We define $\epsilon$ as
\begin{equation}\label{eq:error_dist}
    \epsilon = \max\left(\frac{P_\mathrm{{layer}}}{P_\mathrm{{complete}}},\frac{P_\mathrm{{complete}}}{P_\mathrm{{layer}}}  \right) - 1,
\end{equation}
where $\epsilon = 0$ indicates that $P_\mathrm{{layer}}$ and $P_\mathrm{{complete}}$ are equal. The $\max$ function is necessary to give meaningful values when averaging this quantity. 
We find that to predict the agent's policy with an accuracy of $1\%$, information of $3$-$5$ layers is required.
Results for all action types are shown in \Cref{fig:fig_prob_lay}.

Finally, we compare the agent's policy in a simple scenario to the, in this case, known optimal policy.
Specifically, we take a closer look at the
\emph{Copy} action by evaluating its probability in a class of example diagrams as shown in \Cref{fig:fig_interpret}\,(b).
A phaseless Z-spider is connected to a phaseless X-spider with $n_\mathrm{out}$ additional edges. On $n_\mathrm{extra}$ of those edges, Z-spiders with arbitrary phase are inserted (see inlay). We plot the probability $P_\mathrm{copy}$ of applying the \emph{Copy} action to the edge connecting the phaseless spiders against $n_\mathrm{extra}$ for several $n_\mathrm{out}$. The ideal strategy in this diagram is to apply the \emph{Copy} action if $n_\mathrm{out} - n_\mathrm{extra} \leq 2$ as then multiple \emph{Fuse} actions are enabled, leading to a cumulative positive reward. The agent learns this ideal strategy to good approximation even though it was only trained on random ZX-diagrams and never specifically on diagrams of the type considered here.

\subsection{Scaling}\label{sec:scaling}

\new{Each application of the GNN  requires linear time in the number of nodes and edges in the diagram after which, as currently implemented, one action is applied.  When scaling to larger diagrams, this could be improved: After each evaluation of the GNN, multiple actions can be applied as long as no information has passed through the GNN between the action locations.  Since, in our case, the GNN consists of $6$ message-passing layers, the actions should be separated by at least 6 layers, as introduced in the previous section.
How many actions can be applied simultaneously depends on the connectivity of the ZX-diagram. If the ZX-diagram is extracted from a quantum circuit or measurement-based computation, it typically does not contain long-range connections between spiders and we expect that $\mathcal{O}(n_\mathrm{nodes})$ actions can be applied after each GNN evaluation. Using this approach, the run-time of our RL algorithm scales the same as the \emph{full\textunderscore reduce} algorithm of the \texttt{PyZX} software package\,\cite{kissinger2019}, albeit with a worse prefactor. 
The total number of actions required to simplify a diagram depends heavily on its structure and no clear statement can be made about how it depends on the number of nodes in the diagram.}

\section{Outlook}\label{sec:outlook}

In this work, we have introduced a general scheme for optimizing ZX-diagrams using reinforcement learning with graph neural networks. We showed that the reinforcement learning agent learns non-trivial strategies and generalizes well to diagrams much larger than included in the training set. The presented scheme could be applied to a wide range of problems currently tackled by heuristic and approximate algorithms or simulated annealing.

For example, in \cite{Cam2023} the authors speed up tensor network simulations of quantum circuits by optimizing the graph property treewidth of the corresponding ZX-diagram using simulated annealing, which could straightforwardly be replaced by a reinforcement learning agent.

In \cite{Duncan2020}, a deterministic algorithm for simplification of quantum circuits using ZX-calculus is introduced. The used transformation set is restricted to just two kinds of actions to preserve a special graph property of the ZX-diagram called gFlow, guaranteeing an efficient extraction of a quantum circuit from the optimized diagrams.
Later, a heuristic modification was proposed to reduce the number of two-qubit gates in the resulting circuits\,\cite{Staudacher2022}. 
Meanwhile, also other gFlow preserving rules have been found\,\cite{Mcelvanney2023}. \new{Additionally, the notion of gFLow can be relaxed to the more general Pauli flow which permits additional transformation rules while still allowing efficient circuit extraction\,\cite{simmons_2021}.} However, it is currently unclear when these rules should be applied for the goal of circuit optimization.
In future work, a reinforcement learning agent could be trained including all gFlow \new{or Pauli flow} preserving rules with a reward dependent on the efficiently extracted quantum circuit corresponding to the diagram, thereby taking advantage of new rules and replacing human heuristics with a learned strategy. The agent's reward could, for example, be the total gate, two-qubit gate, or T-gate count.

\new{Finally, the \texttt{PyZX} software package\,\cite{kissinger2019} has been used for quantum circuit equivalence checking using the ZX-calculus\,\cite{Peham2022, Peham_2023}. However, this approach is only guaranteed to work for Clifford circuits due to the limited set of transformation rules of the employed algorithm. Since no circuit needs to be extracted from the optimized ZX-diagram for quantum circuit equivalence checking, an RL agent could use a complete set of transformation rules to potentially overcome this shortcoming.}

During the final preparations of this manuscript, a master thesis using reinforcement learning for quantum circuit compilation with ZX-calculus, albeit using convolutional neural networks, was released\,\cite{Gomez2023}.

\section{Data availability}
Python code of the custom reinforcement learning algorithm using graph neural networks and neural network weights of the trained agents are publicly available on GitHub\,\cite{code}.

\section*{Acknowledgements}
We thank Jonas Landgraf, Jan Olle, and Remmy Zen for fruitful discussions. This research is part of the Munich Quantum Valley, which is supported by the Bavarian state government with funds from the Hightech Agenda Bayern Plus.

\FloatBarrier

\appendix 
\section{Details on ZX-calculus}\label{sec:details_zx}
\begin{figure}
    \centering    
    \includegraphics{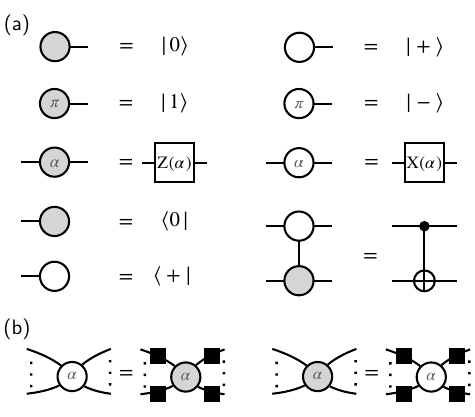}
    \caption{\new{(a) Translation of common quantum gates, states, and (post-selected) measurements into corresponding ZX-diagrams. The translations are true only up to a scalar factor. Square boxes are Z-/X-rotation gates with an angle $\alpha$. (b) By inserting Hadamards (black boxes) on all edges connected to a spider, its color can be changed.}}
    \label{fig:zx_definitions}
\end{figure}

\begin{figure*}
    \centering    
    \includegraphics{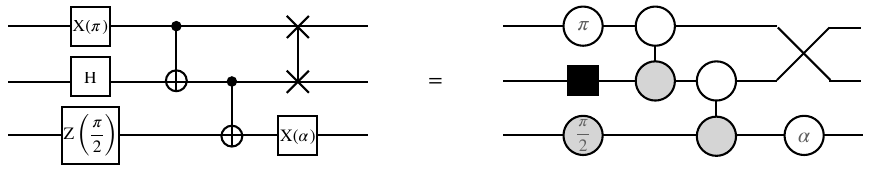}
    \caption{\new{Translation of a quantum circuit into a corresponding ZX-diagram. Single-qubit rotations can be represented by spiders with a single input and output. CNOT gates are represented by two connected spiders. A SWAP gate corresponds to a simple permutation of the edges in the ZX-diagram.}}
    \label{fig:zx_big_circuit}
\end{figure*}

\new{Using the definition of the spiders in \Cref{eq:spider_def}, we can derive representations of common quantum gates, basis states, and measurements in the ZX-calculus [see \Cref{fig:zx_definitions}\,(a)].
Single-qubit Z/X-rotation gates with arbitrary angles can be represented as single Z/X-spiders with one input and one output edge and their phase corresponding to the angle of the rotation gate. Moreover, a CNOT gate can be represented by a phaseless Z-spider connected to a phaseless X-spider. Since this is a complete gate set, it follows that any quantum circuit can be represented as a ZX-diagram. We show an exemplary translation of a quantum circuit into a ZX-diagram in \Cref{fig:zx_big_circuit}. 
However, not every ZX-diagram can easily be represented as a quantum circuit, since ZX-diagrams can also represent states (spiders with only output legs), and post-selected measurements (spiders with only input legs). Similar to a basis change from Z- to X-basis in quantum circuits, the color of spiders can be changed by inserting  Hadamards on all connected edges, as shown in \Cref{fig:zx_definitions}\,(b).
For more details on the ZX-calculus see e.g.\ the review article\,\cite{Wetering2020}.}

\section{Sampled diagrams}\label{sec:details_sampled}

To enable the agent to simplify a wide range of ZX-diagrams, we sample a diverse set of diagrams during training. A typical example is shown in \Cref{fig:fig_result}\,(e). Each new ZX-diagram is constructed with the following steps: First, the number of inputs and outputs is sampled uniformly between $1$ and $3$. \new{Since the RL agent acts locally on the ZX-diagrams without requiring access to their underlying matrix representation, we expect the agent's learned policy to perform similarly well for diagrams with any number of input and output edges.}
Second, the number of initial spiders $n_\mathrm{init}$ is sampled uniformly between $10$ and $15$. The amount of Hadamards is then sampled between $0$ and $\lfloor0.2n_\mathrm{init}\rfloor$. The angles of the initial spiders can be one of $0$, $\pi$, $\pi/2$, and $\alpha$. To determine the angles of the spiders, we uniformly sample a number between $0$ and $1$ for each angle type, reduce the number for $\pi$, $\pi/2$, $\alpha$ by a factor $0.4$ and then normalize the result to a probability distribution from which we sample the angle of each spider. 
We then uniformly sample the expected number of neighbors $n_\mathrm{neigh}$ per spider between $2$ and $4$. From this, we compute the edge probability $p_\mathrm{edge}$ such that when we create each possible edge in the diagram with $p_\mathrm{edge}$ we will have an expected amount of $n_\mathrm{neigh}$ neighbors per spider. We then add each possible edge between all pairs of spiders with probability $p_\mathrm{edge}$ to the diagram.
Finally, we apply the automatic actions that we also apply after each action by the RL agent, i.e.\ removing redundant edges, removing parts of the diagram not connected to any input or output, and applying all possible \emph{Identity} and \emph{Hadamard loop} transformations.
For the performance evaluation of the agent on bigger diagrams we instead sample the number of initial spiders $n_\mathrm{init}$ between $100$ and $150$.
\new{If, instead of randomly sampling ZX diagrams, we had created them by translating quantum circuits, this spider number would correspond to circuits with, for example, up to $75$ single-qubit gates and $75$ two-qubit gates.}

\section{Details on custom PPO algorithm}\label{sec:details_ppo}
\begin{figure*}
    \centering    
    \includegraphics{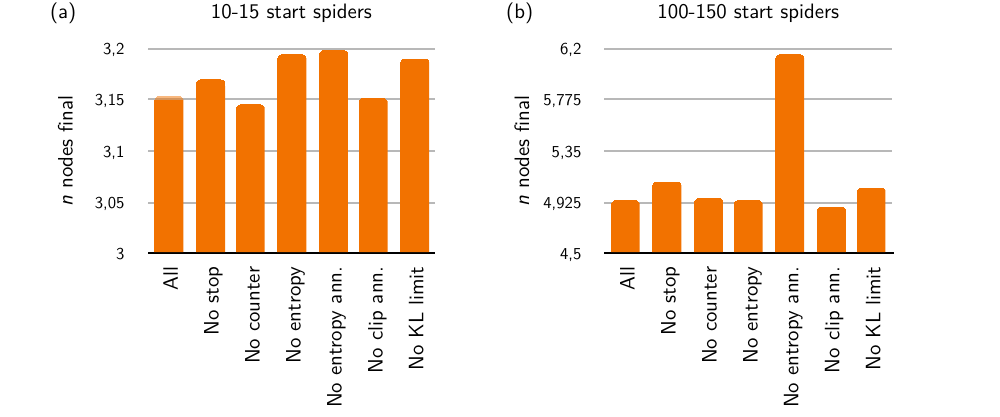}
    \caption{Ablation studies. Average number of nodes left after optimization through an agent trained without a certain feature of the PPO algorithm evaluated over $1000$ ZX-diagrams with $10$-$15$ initial spiders (a) and $100$-$150$ initial spiders (b). Two agents with all features were trained resulting in similar performance (leftmost bars in each diagram). The features that were switched off are the \emph{Stop} action of the agent, the stop counter (see \Cref{sec:details_network}), the entropy bonus $\epsilon$, the annealing of $\epsilon$, the annealing of the clip range $c$, and the early stopping of gradient updates if a Kullback-Leibler divergence of $c_\mathrm{KL}$ is exceeded.}
    \label{fig:fig_ablation}
\end{figure*}

PPO is an actor-critic RL method with a policy network predicting action probabilities and a critic network predicting the so-called advantage of a specific action\,\cite{Schulman2017}. The critic network is only used during training to reduce the variance in gradient update steps. Due to the variable size of our observations and action space, we use a custom implementation of PPO. During the sampling phase of the training, we run $n_\mathrm{env}$ environments in parallel for $n_\mathrm{max}$ steps each. Then, the agent's experiences are randomly split into minibatches of size $n_\mathrm{minibatch}$ which the agent's policy and critic network is then trained on for one gradient step. After the agent is trained on all minibatches, they are reshuffled and another round of training starts for a maximum of $n_\mathrm{train}$ steps. However, if the Kullback-Leibler divergence, estimated as in\,\cite{SchulmannKL}, between the agent's newly trained policy and the policy used in the last sampling phase gets larger than the constant $c_\mathrm{KL}$ we stop the training early and start a new sampling phase. This is not a standard feature of PPO algorithms but has e.g.\ been implemented in \cite{SchulmannModular}. We linearly anneal both the clip range $c$ of the PPO algorithm (as defined in \cite{Schulman2017}) and the entropy coefficient $\epsilon$, which rewards higher entropy of the policy during training leading to more exploration. During training, we clip all gradients to a maximum of $c_\mathrm{absgrad}$ and also clip the norm of the gradients of a minibatch to $c_\mathrm{normgrad}$.
For the gradient updates, we use the ADAM optimizer\,\cite{kingma2014} with a learning rate $\eta$ and exponential moment decay rates $\beta_1$ and $\beta_2$. All parameter values are summarized in \Cref{tab:ppo_paras}, chosen as suggested in \cite{Schulman2017, Andrychowicz2020Whatmatters}, and not further optimized.

We perform ablation studies on some features of the PPO algorithm by switching them off and training a new agent without them. The results are summarized in \Cref{fig:fig_ablation}. 
Entropy annealing has a significant positive impact on the agent's performance when simplifying large diagrams. As a policy with high entropy is more probabilistic, it might need more than the $200$ given steps to fully simplify a large diagram.
All other features don't impact performance significantly. However, we did not optimize the hyperparameters of any of the features which might further increase the performance of the agent.

\begin{table}
\centering
\begin{tabular}{ cc}
\toprule
 parameter & value\\ 
 \midrule
 \multicolumn{2}{c}{PPO} \\
 \midrule
 $n_\mathrm{env}$ & $90$ \\ 
 $n_\mathrm{max}$ & $1000$ \\ 
 $n_\mathrm{minibatch}$ & $3000$ \\ 
 $n_\mathrm{train}$ & $10$ \\ 
 $c_\mathrm{KL}$ & $0.01$ \\ 
 $c$ & $0.2$ \\ 
 $\epsilon$ & $0.1$ \\ 
 $c_\mathrm{absgrad}$ & $100$ \\ 
 $c_\mathrm{normgrad}$ & $0.5$\\ 
 PPO policy loss $\gamma$ & $0.99$\\ 
 PPO policy loss $\lambda$ & $0.9$\\ 
 $\eta$ & $3 * 10^{-4}$\\ 
 $\beta_1$ & $0.9$ \\ 
 $\beta_2$ & $0.999$\\ 
 \midrule
 \multicolumn{2}{c}{GNN} \\
\midrule
message-passing layers & 6 \\
hidden layers $\chi_\mathrm{node}$, $\chi_\mathrm{edge}$ & 1\\
 hidden layers $\chi_\mathrm{stop}$ & 2 \\
all hidden layer dimensions & 128\\
activation functions & tanh \\
 \bottomrule
\end{tabular}
\caption{Parameter values used in the PPO algorithm and GNN. For the definition of $\gamma$ and $\lambda$ see \cite{Schulman2017}.}\label{tab:ppo_paras}
\end{table}

\section{Details on network architecture}\label{sec:details_network}

\begin{figure*}[t]
    \centering
    \includegraphics{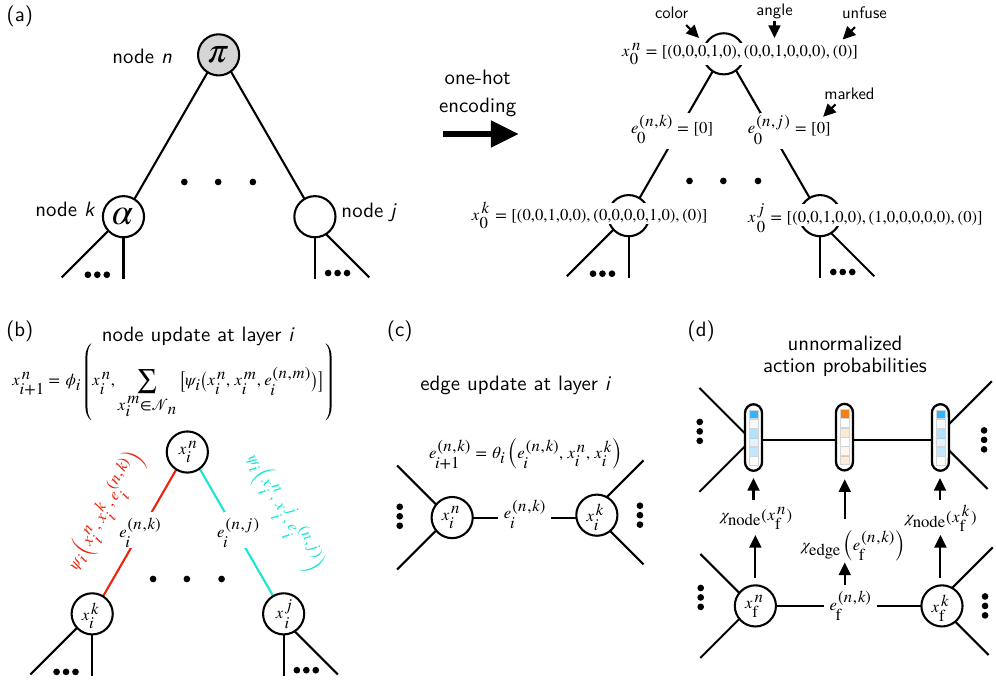}
    \caption{\new{Schematic of the GNN steps. (a) One-hot encoding. Each node is represented by a one-hot encoded five-dimensional color feature, a one-hot encoded six-dimensional angle feature, and an additional number indicating whether the node was previously selected by the \emph{start Unfuse} action. The three features are concatenated to form a single vector. Edges have a one-dimensional feature vector  indicating whether the edge was previously selected by the \emph{Mark edge} action.
    (b) Update of the node feature vector $x_i^n$ at message-passing layer $i$. First, messages for each connected edge are computed using the $\psi_i$ dense neural network layer. Second, the messages are combined by an element-wise sum operation. Finally, the new node feature vector $x_{i+1}^n$ is computed using the $\phi_i$ dense neural network layer which takes the combined messages and the previous node feature vector as an input.
    (c) Update of the edge feature vector $e_i^{(n,k)}$ at message-passing layer $i$. The new edge feature vector $e_{i+1}^{(n,k)}$ is computed using the $\theta_i$ dense neural network layer which takes the feature vectors of the two nodes connected to the edge and the previous edge feature vector as an input. (d) After all message-passing layers are applied, a final multi-layer perceptron $\chi_\mathrm{node}$/$\chi_\mathrm{edge}$ is applied to each node/edge respectively to predict the final unnormalized probabilities of each possible action.}}
    \label{fig:fig_gnn}
\end{figure*}

\begin{table*}
\centering
\begin{tabular}{cccccc}
\toprule
 & \multicolumn{2}{c}{all nodes left} & \multicolumn{2}{c}{$\alpha$ spiders left} & time\,[s] \\
 & $10$-$15$ start &  $100$-$150$ start & $10$-$15$ start & $100$-$150$ start &  $100$-$150$ start\\
\midrule
PyZX + RL & 3.143 & 4.727 & 1.114 & 1.443 & 3.12 \\
RL & 3.148 & 4.934 & 1.122 & 1.484 & 3.91 \\
PyZX + Greedy  & 3.214 & 5.121 & 1.210 & 1.592  &0.79 \\
Sim.\ ann.\ 20000 steps  & 3.379 & 6.212 & 1.356 & 1.871 & 36.7   \\
Greedy  & 3.594 & 10.662 & 1.258 & 3.125 & 122.1 \\
PyZX  & 5.781 & 15.865 & 1.280 & 3.795 & 0.03 \\
\bottomrule
\end{tabular}
\caption{\new{Total number of nodes (left) or number of $\alpha$~spiders (middle) left after optimization with different strategies. The right column is the average run-time to optimize a ZX-diagram with $100$-$150$ start spiders.}}\label{tab:num_results}
\end{table*}

\begin{figure*}
    \centering \includegraphics{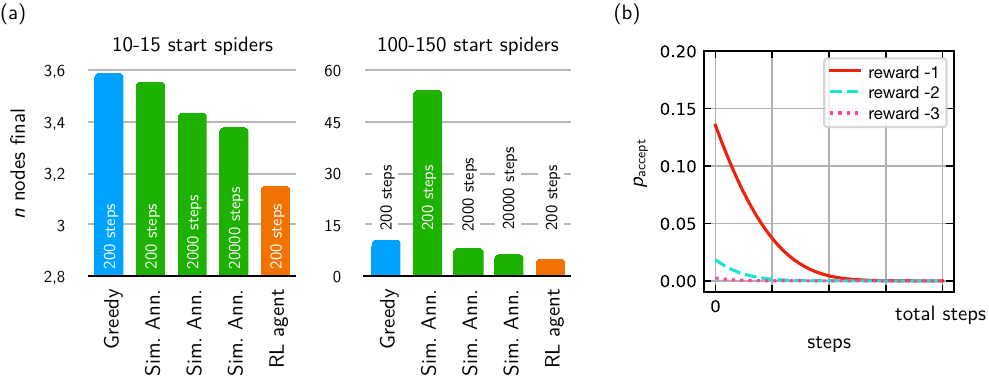}
    \caption{Simulated annealing. (a)~Average number of nodes left after optimization through simulated annealing with start temperature $T_\mathrm{start}=0.5$ evaluated over $1000$ ZX-diagrams with $10$-$15$ starting spiders (left) and $100$-$150$ starting spiders (right). The temperature decay factor $c_\mathrm{ann}$ is chosen as $0.01$/$0.001$/$0.0001$ for $200$/$2000$/$20000$ total steps taken respectively, which results in an acceptance probability of non-greedy actions as shown in (b) for different values of the instantaneous reward of the action.}
    \label{fig:fig_sim_anneal}
\end{figure*}

\begin{figure*}
    \centering \includegraphics{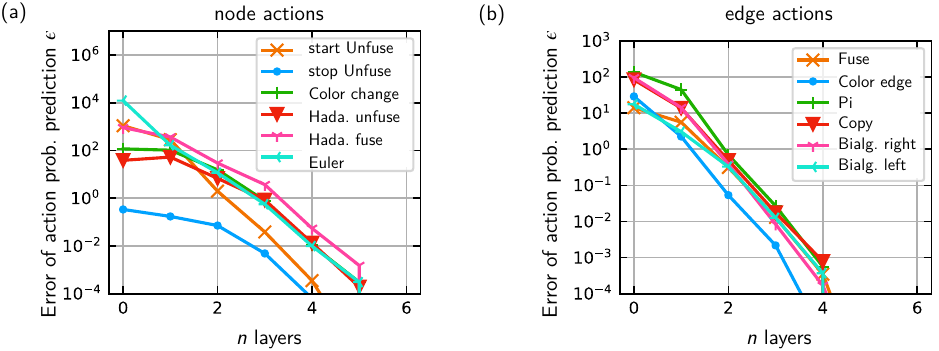}
    \caption{Action probability prediction error $\epsilon$ as defined in \Cref{eq:error_dist} against the number of layers $n$ as defined in \Cref{sec:analysis} for all node actions (a) and edge actions (b).}
    \label{fig:fig_prob_lay}
\end{figure*}

\new{We graphically represent the steps of the GNN network in \Cref{fig:fig_gnn}.}

In the policy network, we use $6$ message-passing layers.
The message functions $\psi_i$, node features computed by $\phi_i$, edge features computed by $\theta_i$, as well as the hidden layers of the final action prediction networks $\chi_\mathrm{node}$, $\chi_\mathrm{edge}$ and $\chi_\mathrm{stop}$, all contain $128$ neurons and use the Tangens hyperbolicus as an activation function. The $\chi_\mathrm{node}$ and $\chi_\mathrm{edge}$ multi-layer perceptrons both have only a single hidden layer, while $\chi_\mathrm{stop}$ has two hidden layers to better learn the more complex global \emph{Stop} action. In addition to the final node/edge states $x_\mathrm{f}^i$/$e_\mathrm{f}^{(n,m)}$, $\chi_\mathrm{node}$/$\chi_\mathrm{edge}$ also get as input an integer number, the stop counter. The stop counter is defined as $\min(20, \mathrm{Steps\ left\ in\ trajectory})$ and tells the agent when a trajectory is about to finish due to the maximum amount of allowed steps being reached. 

The global vector $C$, which is used as part of the input of $\chi_\mathrm{stop}$ contains the number of nodes and the number of edges. Additionally, it holds the number of Z-spiders, X-spiders, Hadamards, spiders with zero/pi/arbitrary angle, and the amount of allowed \emph{Hadamard fuse} and \emph{Euler} actions all normalized by the total spider number and the amount of allowed \emph{Fuse}, \emph{Pi}, \emph{Copy}, \emph{Bialgebra right}, and \emph{Bialgebra left} actions all normalized by the total edge number. Finally, it contains the stop counter and a binary flag, whether the agent has currently selected the \emph{start Unfuse} action. We find that providing the agent global information for predicting the \emph{Stop} action and for predicting the advantage through the critic network is critical to achieving stable training and avoiding exploding gradients as the GNN can otherwise only learn local quantities of the graph.

The critic network has the same network architecture as the network predicting the probability of the \emph{Stop} action but shares no weights with the policy network.

We initialize all trainable parameters of the neural network layers as recommended in \cite{Andrychowicz2020Whatmatters} using an orthogonal initializer with gain $\sqrt{2}$ for all hidden layers, gain $0.01$ for the action prediction networks  $\chi_\mathrm{node}$, $\chi_\mathrm{edge}$ and $\chi_\mathrm{stop}$, and gain $1$ for the final layer of the critic network.

No optimization over the network size or parameters is performed suggesting further possibilities for improving the performance of the RL agent.

\section{Details on simulated annealing}\label{sec:details_sim_ann}

Simulated annealing is a probabilistic algorithm iteratively transforming the ZX-diagrams. At each step, it randomly selects one of all allowed actions. If the immediate reward $r$ of the action is non-negative, the action is applied. If $r$ is negative, the action is only accepted with probability
\begin{equation}
        p_\mathrm{accept} = \exp(r/T),
\end{equation}
 where $T$ is the so-called temperature. $T$ is typically continuously decreased during the optimization process. We choose to exponentially anneal $T$ with the start temperature $T_\mathrm{start}$ at optimization step $n_\mathrm{step}$ according to 
 \begin{equation}
        T = T_\mathrm{start}\exp(-c_\mathrm{ann}n_\mathrm{step}),
    \end{equation}
where $c_\mathrm{ann}$ determines the speed of the temperature decay, as it performs better on the example diagrams than linearly annealing $T$. This may be because the exponential temperature decay leads to a longer nearly greedy phase of the algorithm in the later stages of the optimization.

We further improve the performance of the simulated annealing algorithm by changing the reward structure of the \emph{Unfuse} transformation. Instead of giving $0$ reward when the \emph{start Unfuse} action is selected and $-1$ rewards when the \emph{stop Unfuse} is selected we switch the order of the two rewards. This helps the algorithm to avoid selecting \emph{start Unfuse} in the later, nearly greedy stages of optimization and then getting stuck since it never accepts the negative reward of the \emph{stop Unfuse} action.

We optimize $T_\mathrm{start}$ and $c_\mathrm{ann}$ on two diagrams which the greedy strategy can not optimize well [the diagrams used for \Cref{fig:fig_result}\,(b) and (e)] and then keep them fixed while evaluating the performance of simulated annealing on the same set of diagrams, we evaluated the RL agent on. We find that $T_\mathrm{start} = 0.5$ performs well with $c_\mathrm{ann} = 0.01/0.001/0.0001$ for a maximum of $200/2000/20000$ optimization steps. We also tried $T_\mathrm{start} = 1$ which performed similar to $T_\mathrm{start} = 0.5$ on the example diagrams but considerably worse on average and even higher starting temperatures which even failed to optimize the example diagrams.
As shown in \Cref{fig:fig_sim_anneal}, simulated annealing performs slightly worse on average while needing a lot more optimization steps than the RL agent.

\end{document}